\documentclass[final]{ckm}                 % twocolumn proceedings style

\confname{Workshop on the CKM Unitarity Triangle, IPPP Durham, April 2003}

\def\Eqn#1{Equation~(\ref{#1})}
\def\Fig#1{Figure~\ref{#1}}
\def\Tab#1{Table~\ref{#1}}

\newcommand{\Begeqn}{\begin{equation}}
\newcommand{\Endeqn}{  \end{equation}}

\newcommand{\Begeqnarray}{\begin{eqnarray}}
\newcommand{\Endeqnarray}{  \end{eqnarray}}

\newcommand{\Arraystretch}{1.2}

\newcommand{\Begtabular}[1]{\renewcommand{\arraystretch}{\Arraystretch}
                         \begin{center}\begin{tabular}{#1}\hline\hline}
\newcommand{\Endtabular}{\hline\hline    \end{tabular}\end{center}}

\newlength{\Itemindent}
\def\Itemwidths{
\setlength{\leftmargini}{\Itemindent}
\setlength{\leftmarginii}{\Itemindent}
\setlength{\leftmarginiii}{\Itemindent}
\setlength{\leftmargin}{\Itemindent}
\settowidth{\labelwidth}{\Large$\bullet$}
\setlength{\labelsep}{\leftmargin}
\addtolength{\labelsep}{-\labelwidth}}

\def\Itemspace{
  \topsep    0pt plus 1pt minus 1pt             
  \partopsep 0pt plus 1pt minus 1pt             
  \parskip   0pt plus 1pt minus 1pt 
  \parsep    0pt plus 1pt minus 1pt 
  \itemsep   1pt plus 1pt minus 1pt}

\catcode`\@=11 % @ signs can be used

\newcommand{\Begitem}{\Itemwidths \begin{list}
{\csname\@itemitem\endcsname}
{\ifnum \@itemdepth >4 \@toodeep\else \advance\@itemdepth 1 
 \edef\@itemitem{labelitem\romannumeral\the\@itemdepth}
 \Itemspace \fi}}
\newcommand{\Enditem}{\end{list}}

\catcode`\@=12 % @ signs cannot be used

\def\Vcb{|V_{cb}|}

\def\calB{{\cal B}}
\def\calE{{\cal E}}

\def\calG{{\cal G}}
\def\calM{{\cal M}}
\def\calO{{\cal O}}

\def\calT{{\cal T}}

\def\fbinv{fb$^{-1}$}

\def\nubar{\bar{\nu}}

\def\ellm{\ell^-}

\def\piz{\pi^0}

\def\Dstar{D^*}

\def\Bbar{\bar{B}}

\def\Bstar{B^*}

\def\btosgamma{b \to s \gamma}
\def\BtoXsgamma{B \to X_s \gamma}

\def\BtoXlnu{\Bbar \to X \ell \nubar}
\def\BtoXclnu{\Bbar \to X_c \ell \nubar}
\def\BtoXulnu{\Bbar \to X_u \ell \nubar}

\def\BtoDlnu{\Bbar \to D \ellm \nubar}

\def\BtoDstarlnu{\Bbar \to \Dstar \ellm \nubar}

\def\Jpsi{J/\psi}

\def\fbinv{fb$^{-1}$}

\def\ie{{\it i.e.}}
\def\etal{{\it et al.}}

\def\Lambdabar{\bar{\Lambda}}

\title{CLEO Determinations of $\Vcb$ from Inclusive Moments}

\author{D.G. Cassel}
\address{Laboratory for Elementary-Particle Physics, Cornell University,
Ithaca, NY 14853}

\begin{document}

\begin{abstract}Moments of $\BtoXlnu$ and $\BtoXsgamma$ decays can determine
nonperturbative QCD parameters that relate the semileptonic decay width to $\Vcb$. 
CLEO pioneered measurement of these moments, determined the relevant QCD parameters
from the measured moments, and used these parameters to determine $\Vcb$.\vspace*{-3ex}
\end{abstract}

\maketitle

%% standard LaTeX from here on...

The width $\Gamma_{SL}^c \equiv \Gamma(\BtoXclnu) = \calB(\BtoXclnu) / \tau_B$ for
inclusive semileptonic decay to all  charm states
$X_c$ is related to the CKM matrix element $\Vcb$ by 
$\Gamma_{SL}^c  = {\gamma_c} \Vcb^2.$  Hence, $\Vcb$ can be determined from 
measurements of the lifetime $\tau_B$ and the branching fraction for $\BtoXclnu$
decay, if the  parameter $\gamma_c$ is known.  Unfortunately
$\gamma_c$ is a nonperturbative QCD parameter, and previously theoretical
models had been the only means of estimating $\gamma_c$ \cite{cleoBXclnu,ckmyb}. 
CLEO pioneered determination of $\gamma_c$ from
measurements of energy moments in $\BtoXsgamma$ decays \cite{cleoEgammom} and
measurements of hadronic mass moments in $\BtoXclnu$ decays \cite{cleoMXmom}, and
confirmed the results with measurements of lepton energy moments \cite{cleoElmom} in
$\BtoXlnu$ decays.  These and other moment analyses are reviewed in Ref.~\cite{ckmyb}.

\vspace*{-1ex}
\section{Theoretical Framework}
\vspace*{-1ex}

The theoretical framework for these measurements is  Heavy Quark Effective
Theory, the Operator Product Expansion, and the assumption
of parton-hadron duality in inclusive semileptonic $B$ decays.  These lead to
theoretical predictions that observables in
$\BtoXsgamma$ and
$\BtoXclnu$ decays can be expanded in inverse powers of the $B$ meson mass $M_B$
\cite{MBrefs}.  To order
$1/M_B^3$ the parameter $\gamma_c$ is\\[-3ex]
\Begeqn 
\gamma_c = {G_F^2 M_B^5 \over 192 \pi^3}~ 
\calG(\Lambdabar,\lambda_1,\lambda_2|
\rho_1,\rho_2,\calT_1,\calT_2,\calT_3,\calT_4) \label{eq:gammac}
\Endeqn
where $\calG$ is a polynomial in $1/M_B$ and 
$\Lambdabar$, $\lambda_1$, $\lambda_2$,
$\rho_1$, $\rho_2$, $\calT_1$, $\calT_2$, $\calT_3$, and $\calT_4$ are nonperturbative
QCD parameters.  Of these parameters: $\Lambdabar$ appears in all orders above 0 in
$1/M_B$; $\lambda_1$ and $\lambda_2$ first appear in second order; and the others first
appear in third order.  Some of the coefficients of the
polynomial involve expansions in $\alpha_S$, which are carried out to order
$\beta_0\alpha_S^2$.  The power series and the results depend on the 
renormalization scheme used in the theoretical calculations; the calculations used in
these analyses were done in the $\overline{MS}$ scheme.  

Physical interpretation of the parameters $\Lambdabar$, $\lambda_1$, and
$\lambda_2$, comes from the relationship between the
$b$ quark mass $m_b$ and the $B$ and $B^*$ meson masses,
$M_B$ and $M_{B^*}$:
$M_B~ = m_b + \Lambdabar - (\lambda_1 + 3\lambda_2) / (2m_b) + \ldots$ and 
$M_{\Bstar} =  m_b + \Lambdabar - (\lambda_1 - \lambda_2) / (2m_b) + \ldots$~ .
Intuitively, $\Lambdabar$ is the energy of the light quark and gluon
degrees of freedom,
$-\lambda_1$ is the average of the square of the $b$ quark momentum, and 
$\lambda_2/m_b$ is the hyperfine interaction of the $b$ quark and light degrees of
freedom.  Using these expressions, we determine $\lambda_2$ from 
$M_{\Bstar} - M_B \approx 46$ MeV/$c^2$.

There are similar expressions -- involving the same nonperturbative QCD parameters
-- for the moments
$\langle (M_X^2 - \bar{M}_D^2) \rangle/M_B^2$ of the hadronic mass ($M_X$)
spectrum in $\BtoXclnu$ decay and 
$\langle E_\gamma \rangle/M_B$ of the photon energy ($E_\gamma$) spectrum in 
$\BtoXsgamma$ decay.  (Here
$\bar{M}_D = 0.25 M_D + 0.75 M_{\Dstar}$, the spin-averaged $D$ meson mass.) The
coefficients $\calM_n$ and $\calE_n$ of the polynomials for these moments depend on
the lepton momentum range measured in $\BtoXclnu$ decays and the energy range
measured in $\BtoXsgamma$ decays, respectively \cite{Adam,Falketal}.  In the
$M_X^2$ moments $\Lambdabar$ again appears in all orders of $1/M_B$, while $\lambda_1$
and $\lambda_2$ appear in second order.  The latter two parameters do not appear up to
third order in the expansion for $\langle E_\gamma \rangle/M_B$.

We obtained $\gamma_c$ from \Eqn{eq:gammac},
by determining $\Lambdabar$ and $\lambda_1$ from measurements of
$\langle (M_X^2 - \bar{M}_D^2) \rangle$ and $\langle E_\gamma \rangle$
after: determining $\lambda_2$ from $M_{\Bstar} - M_{B}$ and estimating
$\rho_1,\rho_2,\calT_1,\calT_2,\calT_3,\calT_4$ to be at most $(0.5~{\rm
GeV})^3$ from dimensional considerations.  We also measured the second moments of
these distributions, but do not use them to determine $\Lambdabar$ and $\lambda_1$ due
to the current state of theoretical uncertainties.

\begin{figure*}[tb]
\hfill\includegraphics[width=0.4\hsize]{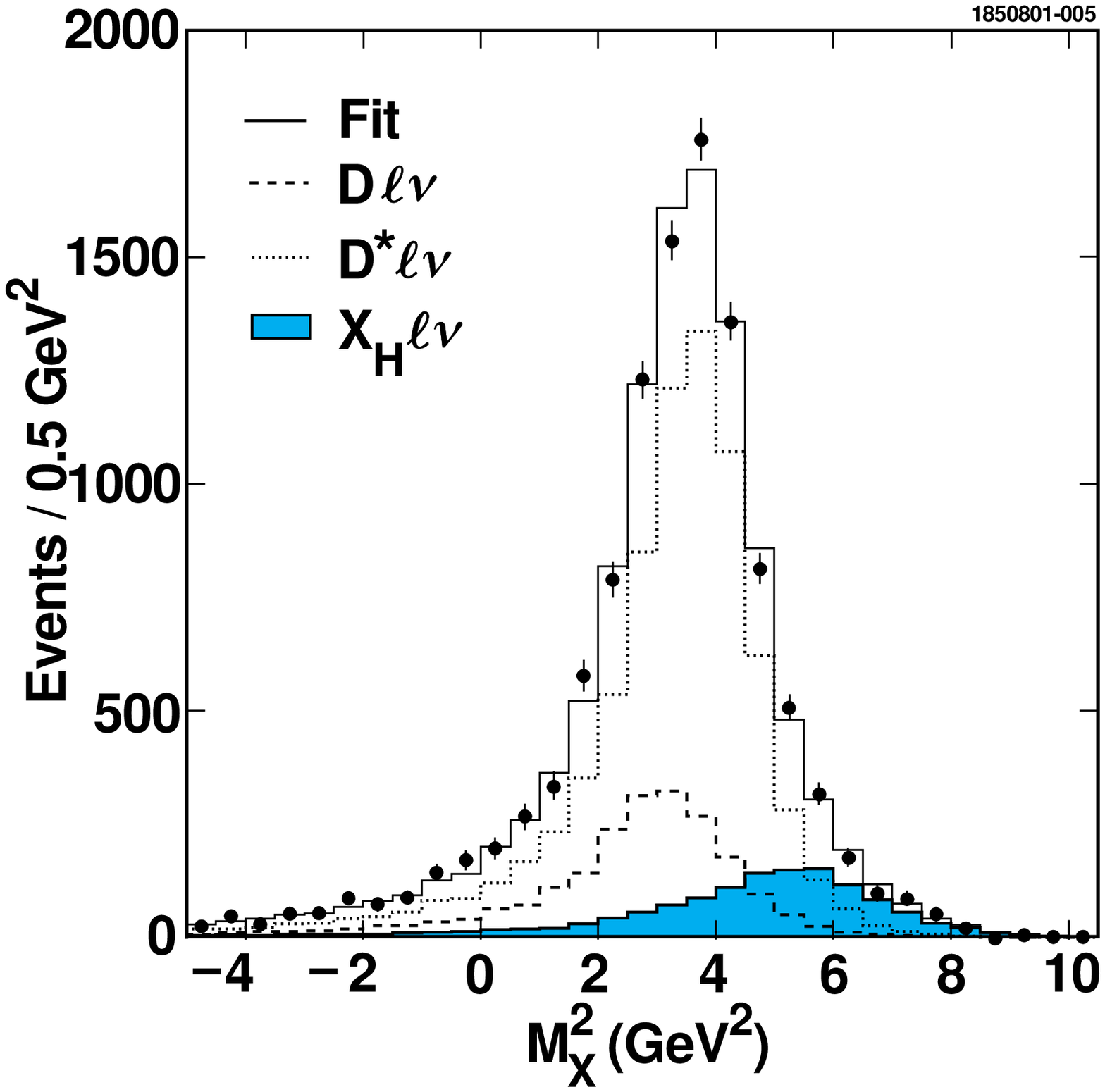}
\hfill\includegraphics[width=0.342\hsize]{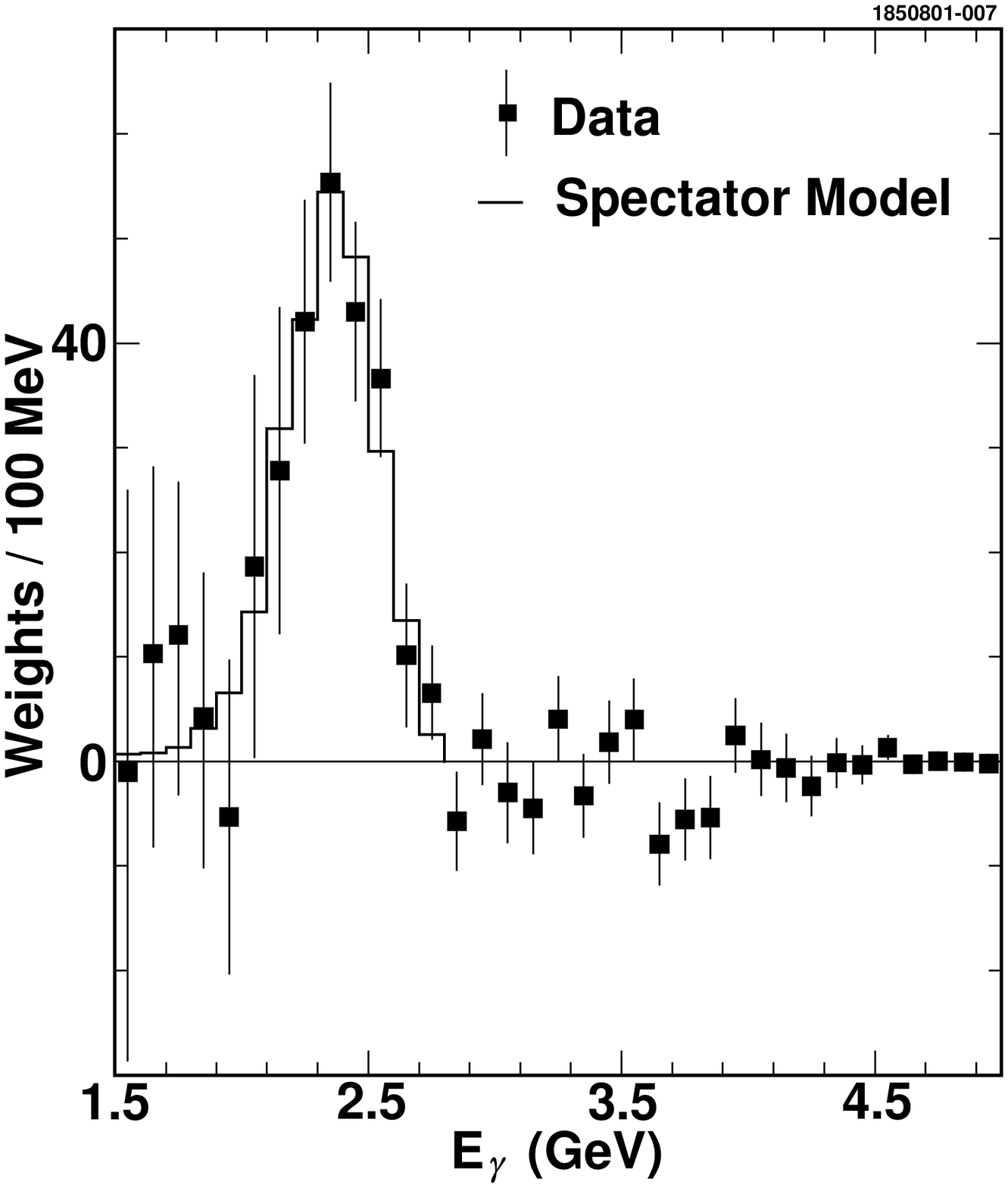}\hfill\hfill
\caption{(left) Number of events versus $M_X^2$, the square of the hadronic mass in
$\BtoXclnu$ decay. (right) Weights plotted versus the photon energy $E_\gamma$ in
$\btosgamma$ events.}
\label{fig:mxsqegamma}
\end{figure*}

\vspace*{-1ex}
\section{Measuring the $M_X^2$ and $E_\gamma$ Moments}
\vspace*{-1ex}

We measured the $M_X^2$ moments using 3.2 \fbinv\ of $\Upsilon(4S)$ data and 1.6
\fbinv\ of continuum data of continuum data collected below the $B\bar{B}$ threshold. 
These data were accumulated using the CLEO~II detector \cite{cleoMXmom}.  
The continuum events were used to estimate backgrounds from continuum data in the
$\Upsilon(4S)$ data sample. Calculation of the hadronic mass
$M_X$ started with reconstruction of the  neutrino in events with a single lepton by
ascribing the missing energy and momentum to the neutrino.  We then used\break 
$M_X^2 \cong M_B^2 + M_{\ell\nu}^2 - 2 E_B E_{\ell\nu}$ where $M_{\ell\nu}$ and
$E_{\ell\nu}$ are the invariant mass and the energy of the $\ell\nu$ system,
respectively.  (This expression is obtained by setting 
$\cos\theta_{B-\ell\nu} = 0$, where $\theta_{B-\ell\nu}$ is the unmeasurable 
angle between the momenta of the $B$ and the $\ell\nu$ system.)  
Neutrino energy and momentum resolution, and neglect of the modest term involving
$\cos\theta_{B-\ell\nu}$ result in non-negligible width for the $M_X^2$ distributions
of $\BtoDlnu$ and $\BtoDstarlnu$ decays.  \Fig{fig:mxsqegamma} illustrates the $M_X^2$
distribution obtained in this analysis.

Moments were obtained from fits to the spectrum that include contributions from
$D\ell\nubar$, $D^*\ell\nubar$, and
$X_H\ell\nu$, where $X_H$ represents all higher mass resonant and non-resonant charm
states.  The moments determined in this manner are not very sensitive to the $M_X^2$
distributions assumed for the $X_H$ states and the modest sensitivity is included
in the systematic error.  The measured moments are:
\Begeqnarray
\langle (M_X^2 - \bar{M}_D^2) \rangle\;\,  &=& 0.251 \pm 0.023
\pm 0.062\; {\rm GeV}^2 \nonumber \\
\hspace*{-\mathindent}\langle (M_X^2 - \langle M_X^2 \rangle)^2 \rangle  &=& 
0.639 \pm 0.056 \pm 0.178\, {\rm GeV}^4 \nonumber
\Endeqnarray
We measured $E_\gamma$ moments using 9.1 \fbinv\ of $\Upsilon(4S)$ data and 4.4
\fbinv\ of continuum data collected below the $B\bar{B}$ threshold.  These data were
accumulated using the CLEO~II and CLEO~II.V detector configurations \cite{cleoEgammom}.
The analysis began with a search for an isolated $\gamma$ with
$2.0 < E_\gamma < 2.7$ GeV.  In this energy range, backgrounds are about a factor of
100 above the signal.  Most of these backgrounds are $\gamma$s from
Initial State Radiation or photons from the decay of $\piz$s in continuum events. 
Substantial background reduction is achieved with requirements on event shapes and
energies  in cones relative to $\mathbf{p}_\gamma$, or with pseudoreconstruction of
the $X_s$ state, or by requiring the presence of a lepton in the event.  For each
$\gamma$ candidate, all information was combined into a single weight that ranged 
between 0.0 for continuum events and 1.0  for $\BtoXsgamma$ events.  Using the
continuum data to subtract backgrounds from continuum events in the $\Upsilon(4S)$
data was crucial for this analysis.
The resulting weight distribution is illustrated in \Fig{fig:mxsqegamma}.  The
$E_\gamma$ moments obtained from this weight distribution are:
\Begeqnarray
\langle E_\gamma \rangle~~~~~~~ &=& 2.346\;\, \pm 0.032\;\, \pm 0.011~~\, 
{\rm GeV} \nonumber \\
\langle (E_\gamma -\langle E_\gamma \rangle)^2 \rangle &=& 
0.0226 \pm 0.0066 \pm 0.0020\; {\rm GeV}^2 \nonumber 
\Endeqnarray

\begin{figure}[htb]
\hbox to\hsize{\hss
\includegraphics[width=0.89\hsize]{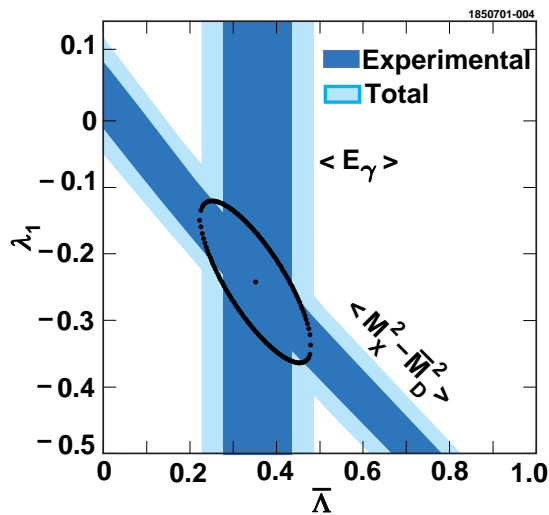}
\hss}
\caption{Bands of the measurements of hadronic mass moments and $E_\gamma$ 
moments plotted in the $\lambda_1$--$\Lambdabar$ plane.}
\label{fig:lambda1lambadabar}

\end{figure}

The measured $E_\gamma$ and $M_X^2$ moments are plotted in the
$\Lambdabar$--$\lambda_1$ plane in \Fig{fig:lambda1lambadabar}. The intersection of
these moments yields correlated values of $\Lambdabar$ and $\lambda_1$,\newpage
\vspace*{-9ex}
\Begeqnarray
\Lambdabar~ &=&\, +0.35\; \pm 0.07~\, \pm 0.10~~~\! \mathrm{GeV~~ and}
                                                   \label{eq:Lambdabar}\\
\lambda_1   &=& -0.238 \pm 0.071 \pm 0.078~ \mathrm{GeV}^2, \label{eq:lambda1}
\Endeqnarray
where the first errors are from the uncertainties in the moment measurements and the
second errors are from the theoretical uncertainties, particularly the uncertainties
in the values of the parameters $\rho_1$, $\rho_2$, $\calT_1$, $\calT_2$, $\calT_3$,
and $\calT_4$, which are not measured.

\vspace*{-1ex}
\section{Determining $\Vcb$ from $M_X^2$ and $E_\gamma$ Moments}
\vspace*{-1ex}

To compute $\Gamma_{SL}^c$, we used:
$\calB(\BtoXclnu) = (10.39 \pm 0.46)\% $ \cite{semi},
$\tau_{B^\pm} = (1.548 \pm 0.032)$ ps \cite{PDG2000},
$\tau_{B^0} = (1.653 \pm 0.028)$ ps \cite{PDG2000},
$f_{+-}/f_{00} = 1.04 \pm 0.08$ \cite{Sylvia},
giving $\Gamma_{SL}^c = (0.427 \pm 0.020) \times 10^{-10}$ MeV.
Then $\Gamma_{SL}^c = \gamma_c\Vcb^2$ and \Eqn{eq:gammac} for $\gamma_c$ then yielded,
\Begeqn
\Vcb = (40.4 \pm 0.9 \pm 0.5 \pm 0.8) \times 10^{-3}, \label{eq:VcbMXEgamma}
\Endeqn
where the first error is from the experimental determination of $\Gamma^c_{SL}$, the
second from the measurement of $\Lambdabar$ and $\lambda_1$, and the
third from theoretical uncertainties, \ie, from
$\alpha_s$ scale uncertainties and ignoring the
$\calO(1/M_B^3)$ terms which contain the estimated parameters $\rho_1$, $\rho_2$,
$\calT_1$, $\calT_2$, $\calT_3$, and $\calT_4$.

Note that -- even with direct measurement of these nonperturbative QCD parameters --
the residual theoretical uncertainty is comparable to the experimental errors!

\vspace*{-1ex}
\section{Determining $\Vcb$ from Moments of the Lepton Spectra in
$\BtoXlnu$}
\vspace*{-1ex}
\label{sec:R0R1}

Moments of the $E_\ell$ spectrum in inclusive $\BtoXlnu$ decay  can also be used to
determine {$\Lambdabar$} and {$\lambda_1$}.
We define:
\[
{R_0 = { \int\limits_{1.7~{\rm GeV}} 
         {d\Gamma_{SL}\over dE_\ell} dE_\ell \over
        \int\limits_{ 1.5~{\rm GeV}} {d\Gamma_{SL}\over dE_\ell} dE_\ell }}
\hspace*{2.3em}
{R_1 = { \int\limits_{ 1.5~{\rm GeV}} E_\ell\, 
         {d\Gamma_{SL}\over dE_\ell} dE_\ell \over
        \int\limits_{ 1.5~{\rm GeV}}~~~\, {d\Gamma_{SL}\over dE_\ell} dE_\ell }}
\]
In order to avoid the necessity of removing
$\BtoXulnu$ decays from our data, we include both $\Gamma^c_{SL}$ and $\Gamma^u_{SL}$
in $\Gamma_{SL}$.  The moments $R_0$ and $R_1$ for $\BtoXlnu$ decay can be expressed
in terms of expansions in $\alpha_S$ and $\bar{M}_B$ (spin-averaged $B$ mass) involving
{$\Lambdabar$} and {$\lambda_1$}.  These expansions have been calculated to
$\calO(1/\bar{M}_B^3)$
\cite{gklw}. Determining $\Lambdabar$ and $\lambda_1$ from these moments provides an
important check of theory, particularly the  importance of neglected higher order
terms in the theoretical $E_\ell$, $M_X^2$, and $E_\gamma$ moments. 

\begin{figure}[htb]
\hbox to\hsize{\hss
\includegraphics[width=0.90\hsize]{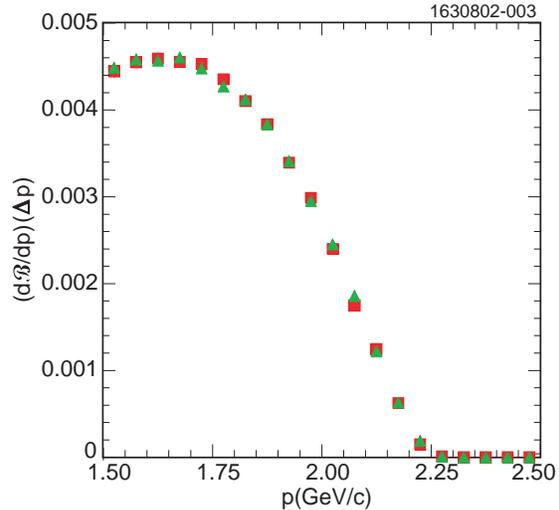}
\hss}
\caption{Momentum spectra in the $B$ meson rest frame for electrons (triangles) and
muons (squares).  The quantity $d{\cal B}/dp$ represents the differential semileptonic
branching fraction in the bin $\Delta p$, divided by the number of $B$ mesons in the
sample and $\Delta p$.}\label{fig:pepmu}
\end{figure}

We measured these moments with the same data sample that we use in the measurement of
the hadronic mass moments. The principal experimental challenges are identifying
leptons and eliminating leptons from sources other than $\BtoXlnu$ decay.  Other
sources include $\Jpsi$ decay, $e^\pm$ from $\piz$ Dalitz decay, $\gamma$
conversions, secondary leptons from $b \to c$ decay followed by $c \to s\ell\nu$
decay, leptons from continuum events, and hadrons misidentified as leptons (this
background is much more significant for muons than for electrons).  After cuts to
reduce leptons from these sources and subtraction of estimated residual yields, we
subtract the yield from our continuum data.   The resulting electron and muon momentum
spectra are illustrated in \Fig{fig:pepmu}.  \Tab{tab:R0R1} gives the
$R_0$ and $R_1$ values obtained from electron, muon, and combined lepton data samples. 
The results for electrons and muons are obviously very consistent.  The dominant
systematic errors are from the secondary lepton contribution, lepton
identification, electroweak radiative corrections, and the uncertainty in the absolute
momentum scale.\\[-4ex]

\begin{table}[h]\caption{Measured values of $R_0$ and $R_1$ from $e$ and
$\mu$ data and for the weighted average of the two ($\ell$). The errors are
statistical and systematic in that order.}\label{tab:R0R1}
\begin{center}
\Begtabular{lccc}
$R_0$ & $e$    & $0.6184 \pm 0.0016 \pm 0.0017$ \\
$R_0$ & $\mu$  & $0.6189 \pm 0.0023 \pm 0.0020$ \\ \hline
$R_0$ & $\ell$ & $0.6187 \pm 0.0014 \pm 0.0016$ \\ \hline\hline
$R_1$ & $e$    & $1.7817 \pm 0.0008 \pm 0.0010$ & GeV \\
$R_1$ & $\mu$  & $1.7802 \pm 0.0011 \pm 0.0011$ & GeV \\ \hline 
$R_1$ & $\ell$ & $1.7810 \pm 0.0007 \pm 0.0009$ & GeV \\
\Endtabular
\end{center}
\end{table}

The values of $\Lambdabar$ and $\lambda_1$ obtained from the lepton
energy moments are,\newpage
\vspace*{-9ex}
\Begeqnarray
\Lambdabar~ &=&\, +0.39 \pm 0.03 \pm 0.06 \pm 0.12~~ \mathrm{GeV~~ and}\\
\lambda_1   &=& -0.25 \pm 0.02 \pm 0.05 \pm 0.14~~ \mathrm{GeV}^2,
\Endeqnarray
where the errors are statistical, systematic, and theory.  The uncertainties in
the  $1/\bar{M}_B^3$ terms dominate the theoretical errors.  (Note that the
theoretical uncertainties are larger than the experimental errors in this analysis!) 
These results are in excellent agreement with the values from the 
$E_\gamma$--$M_X^2$ moments given in Equations~(\ref{eq:Lambdabar}) and
(\ref{eq:lambda1}).

\begin{figure}[htb]
\hbox to\hsize{\hss
\includegraphics[width=0.9\hsize]{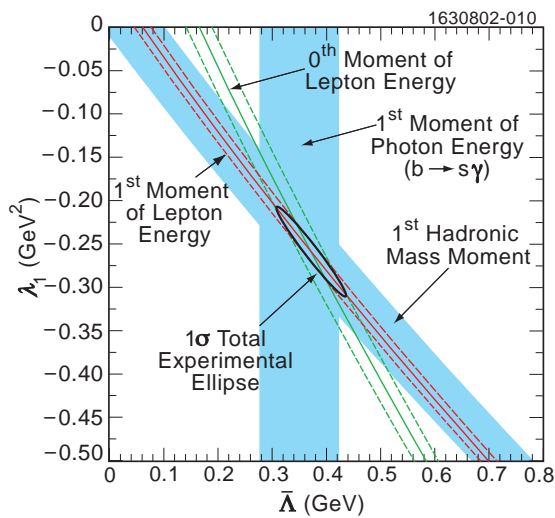}
\hss}
\caption{The electron energy moments $R_0$ and $R_1$ plotted in the 
$\Lambdabar$--$\lambda_1$ plane, along with the $M_X^2$ and $E_\gamma$
moments.  The errors shown are $\pm 1\sigma$ total experimental
errors (statistical plus systematic).}\label{fig:allmoments}
\end{figure}

The agreement between the $R_0$--$R_1$ and $E_\gamma$--$M_X^2$ moment analyses is
illustrated in \Fig{fig:allmoments} where all four measured moments with their total
experimental errors are plotted in the
$\Lambdabar$--$\lambda_1$ plane.    Due to the theoretical uncertainties in the
relationships between moments and 
$\Lambdabar$ and $\lambda_1$ we do not make an overall fit to the two analyses.
There is little correlation among these measurements, so the consistency of the
{$\Lambdabar$} and {$\lambda_1$} values from $E_\gamma$--$M_X^2$ moments with those
from  from $E_\ell$ moments increases confidence in the theories.

The value of $\Vcb$ obtained from these $E_\ell$ moments is
\Begeqn
\Vcb = (40.8 \pm 0.5 \pm 0.4 \pm 0.9) \times 10^{-3}, 
\Endeqn
where the first error is from the experimental determination of $\Gamma^c_{SL}$, the
second from the measurement of $\Lambdabar$ and $\lambda_1$, and the
third from the theoretical uncertainties described following \Eqn{eq:VcbMXEgamma}.
Of course, since the values of $\Lambdabar$ and $\lambda_1$ from this analysis are in
excellent agreement with the corresponding results from the $E_\gamma$--$M_X^2$
analysis, this value of $\Vcb$ agrees very well with the value given in
\Eqn{eq:VcbMXEgamma}.
																			
%\vspace*{-2ex}
\section{Summary}
\vspace*{-1ex}

CLEO determined $\Vcb$ from two different moment analyses.
The result of the $E\gamma$--$M_X^2$ analysis is 
$\Vcb = (40.4 \pm 0.9 \pm 0.5 \pm 0.8) \times 10^{-3}$ while the result of the
$R_0$--$R_1$ analysis is 
$\Vcb = (40.8 \pm 0.5 \pm 0.4 \pm 0.9) \times 10^{-3}$.  In each case, the first
error is the uncertainty due to the uncertainty in the measured
semileptonic decay width $\Gamma(\BtoXclnu)$, the second is from the determination of
$\Lambdabar$ and
$\lambda_1$ from  moments, and the
third is an estimate of the theoretical uncertainty due to the estimated range of
values of additional nonperturbative QCD parameters that are not determined in the
analyses.   The two results are clearly in excellent agreement.  
 
\vspace*{-1ex}

\section*{Acknowledgements}

\vspace*{-1ex}

I am delighted to acknowledge the contributions of my CLEO and
CESR colleagues which led to the results described in this contribution.  I
appreciate the support of the National Science Foundation for the CESR/CLEO
program and the NSF and U.S. Department of Energy for support of my CLEO
collaborators.  Finally, I want to thank the organizers and their staff who have
worked so hard to make this workshop pleasant and successful. 

% A useful Journal macro
\def\Journal#1#2#3#4{{#1} {\bf #2}, #3 (#4)}
\def\Report#1#2#3{#1 Report No.\ #2 (#3)}
\def\Journalmore#1#2#3{{\bf #1}, #2 (#3)}

\def\CLEO{CLEO Collaboration}
% Some useful journal names
\def\EPJC{{\it E.\ Phys J.} C}
\def\NCA{\it Nuovo Cimento}
\def\NIM{\it Nucl.\ Instrum.\ Methods}
\def\NIMA{{\it Nucl.\ Instrum.\ Methods} A}
\def\NPB{{\it Nucl.\ Phys.} B}
\def\PLB{{\it Phys.\ Lett.} B}
\def\PRL{\it Phys.\ Rev.\ Lett.\ }
\def\PRD{{\it Phys.\  Rev.} D}
\def\ZPC{{\it Z.\ Phys.} C}


\begin{thebibliography}{99}
\bibitem{MBrefs}References to the theoretical literature can be found in Refs.
\cite{cleoEgammom,cleoMXmom,cleoElmom}
\end{thebibliography}
\end{document}